\documentclass[conference]{IEEEtran}
\IEEEoverridecommandlockouts
\usepackage{cite}
\usepackage{amsmath,amssymb,amsfonts}
\usepackage{algorithmic}
\usepackage{graphicx}
\usepackage{textcomp}
\usepackage{multirow}
\usepackage{hyperref}
\usepackage{xcolor}
\def\BibTeX{{\rm B\kern-.05em{\sc i\kern-.025em b}\kern-.08em
    T\kern-.1667em\lower.7ex\hbox{E}\kern-.125emX}}

\newcommand{\dataname}{\textit{TweetPap}}
\begin{document}

\title{TweetPap: A Dataset to Study the Social Media Discourse of Scientific Papers}

\author{\IEEEauthorblockN{Naman Jain and Mayank Singh}
\IEEEauthorblockA{Department of Computer Science and Engineering \\
\textit{Indian Institute of Technology Gandhinagar}\\
Gandhinagar, India \\
singh.mayank@iitgn.ac.in}}
\maketitle

\begin{abstract}
Nowadays, researchers have moved to platforms like Twitter to spread information about their ideas and empirical evidence. Recent studies have shown that social media affects the scientific impact of a paper. However,  these studies only utilize the tweet counts to represent Twitter activity. In this paper, we propose \textit{\dataname{}}, a large-scale dataset that introduces temporal information of citation/tweets and the metadata of the tweets to quantify and understand the discourse of scientific papers on social media. The dataset is publicly available at \url{https://github.com/lingo-iitgn/TweetPap}.
\end{abstract}

\section{Introduction}

As digitization led to the availability of a vast volume of publications, measuring research impact became important in order to analyze a publication's performance and optimizing search engines to retrieve relevant articles \cite{b1}. Research institutes also use research impact to assess the performance of faculty members, make grant allocations, set policies, and recruit new members \cite{b3}. The eagerness to improve research impact led researchers to ``push'' their work on social media platforms like Twitter for greater reach\cite{b4,b5}. Existing studies have attempted to capture the discourse of scientific knowledge through social media platforms by analyzing the number of mentions of a publication on Twitter in small-scale datasets created manually or using surveys and analytics platforms \cite{b6, b2, b3, b4} whereas large-scale datasets lack analysis with citation data \cite{b5}.



We propose a novel data collection scheme to broaden the study of quantifying the effect of social media activity on the research impact of scholarly documents and present \textit{\dataname{}} a large-scale dataset consisting of Twitter activity of arXiv papers created using this scheme. Other than the tweet count and citation count of papers, as introduced in previous studies, \textit{\dataname{}} also provides access to yearly citations of a paper, yearly retweets of a particular tweet mentioning a paper, likes on each tweet, and a list of embedded URLs for understanding the discourse of scientific knowledge.


\begin{table}[!tbh]
    \centering
    \caption{Statistics of the \dataname{} Dataset}
    \label{fig:stats}
    \begin{tabular}{c|c}
    \hline 
    \textbf{Year range} & 2010--2019 \\ 
     \textbf{Tweets}  &  367,124 \\
       \textbf{Papers} & 125,521 \\  
       \textbf{Links} & 418,494 \\
       \textbf{Users} & 16,098 \\ \hline 
       \end{tabular}
\end{table}

\section{Data Collection Scheme}
To assess the diffusion of scientific papers on social media platforms, we need a comprehensive dataset with bibliographic information and the corresponding social media activity. Due to open-access permissions and public availability of citation data, \textit{\dataname{}}  is compiled from the papers in arXiv. We collected the relevant data by searching the keyword \textit{``arXiv''} in Twitter's historical database. The tool\footnote{github.com/Jefferson-Henrique/GetOldTweets-python} retrieves all the tweets mentioning the keyword \textit{``arXiv''} in its metadata (text, username and comments). We collected tweets from a time range of 10 years (2010--2019) and mapped the papers to the tweets using arXiv identifiers (ID). We chose arXiv ID because it can be easily retrieved using regular expressions, whereas paper title matching may result in high computational cost and false matches due to variation in punctuation, grammar, and presence of incomplete titles\footnote{Only 16,435 tweets were retrieved using title matching}. Apart from the tweet's text, the arXiv IDs can also be retrieved from the links mentioned in the tweet in many cases. A major challenge in extracting IDs from the links in the tweets is that most of the links are shortened (discussed more in \ref{subsec:links}).  To extract arXiv IDs from such links, we un-shortened\footnote{github.com/SMAPPNYU/urlExpander} all the links from the tweets which do not have an arXiv ID in the non-hyperlink text of the tweet. Finally, using the regular expression ``[0-9]\{4\}.[0-9]\{4\}[0-9]?'' (based on the arXiv guidelines\footnote{https://arXiv.org/help/arXiv\_identifier}), we extracted the arXiv ID from unshortened links. We do not consider the tweets without a retrievable arXiv ID from links or text. Further, we leverage the Semantic Scholar Corpus~\cite{b8} to extract the yearly citation information of arXiv papers. From the arXiv IDs in the tweets, we extracted yearly citation information of 125,521 papers occurring in 367,124 tweets. Since the same paper can be mentioned in multiple tweets, we present a cumulative count of retweet/like from all the tweets corresponding to that paper. In the end, we obtain a dataset with statistics and attributes defined in Table \ref{fig:stats} and Figure \ref{fig:attributes} respectively, along with yearly citations and tweet information such as text and yearly retweets data (pulled from Twitter API\footnote{https://python-twitter.readthedocs.io/en/latest/}).

\begin{figure}
    \centering
    \resizebox{\hsize}{!}{
    \begin{tabular}{ll}
       ``\textbf{arXiv ID}: $<$id$>$''  & ``\textbf{Citation}: $<$map[year$\rightarrow$int]$>$'' \\
    ``\textbf{Publication Year}: $<$int$>$'' & ``\textbf{Retweet}: $<$map[year$\rightarrow$int]$>$'' \\
    ``\textbf{Tweet IDs}: $<$list[int]$>$'' & ``\textbf{Likes}: $<$int$>$''\\
    ``\textbf{Links}: $<$list[string]$>$'' & ``\textbf{Users}: $<$list[string]$>$'' \\
    \end{tabular}}
    \caption{Attributes of \textit{\dataname{}} alongwith data types}
    \label{fig:attributes}
\end{figure}

\section{Dataset Analysis}

This section describes the analysis of \textit{\dataname{}}, focusing on features showcasing the significance of this dataset.


\subsection{Distribution of attributes}
\label{sec:dist}
Figure \ref{fig:urluser} (a) shows the frequency distribution of tweets and papers based on the year of release. Contrary to our expectation, we found a significantly low number of tweets between 2015--2018. \textit{\dataname{}} confirms positive correlation between Twitter activity and citation count. For example, the average citation count increases from 7.88 to 21.2 as the number of tweets per paper increases from one to five. 


\subsection{Temporal peaks of citations and retweets}
\label{sec:temp}
\begin{figure}
\centering
\resizebox{\hsize}{!}{
\begin{tabular}{cc}
    \includegraphics[height=4cm]{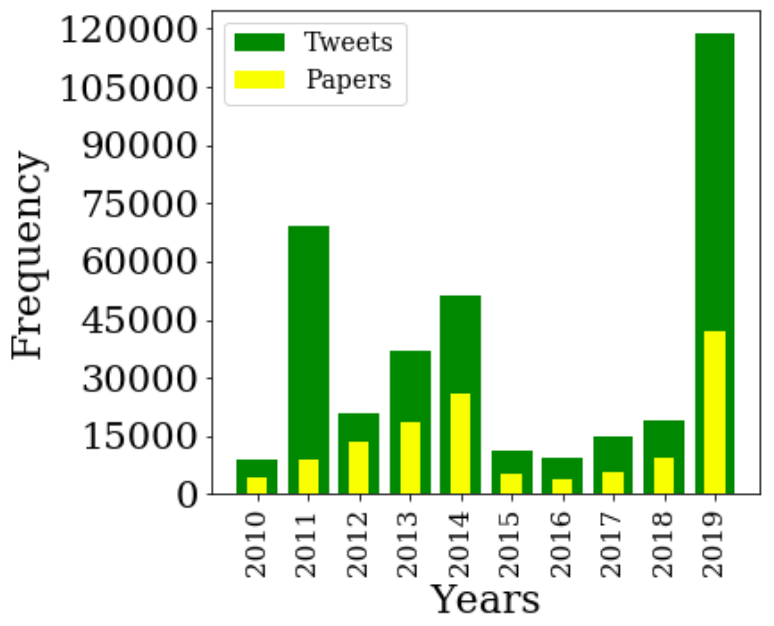}
    &
    \includegraphics[height=4cm]{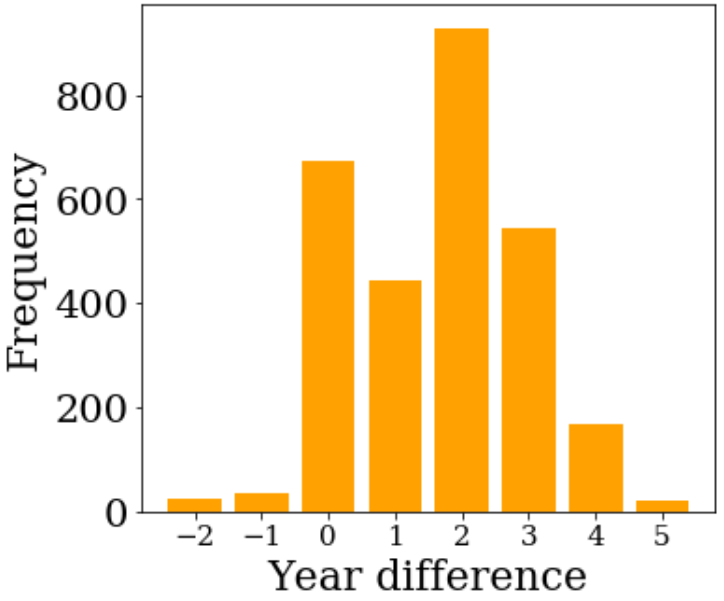}\\
    (a)&(b)
\end{tabular}}
\caption{(a) Frequency distribution of tweets and papers over the years \\(b) Peak difference distribution}
\label{fig:urluser}
\end{figure}

We extract the peak years for both of these attributes by utilizing the yearly citations and retweet data.  A peak year refers to the year in which the corresponding attribute was the highest for a paper. In the case of multiple peak years (years with similar maximum frequency), we consider the first peak. Figure \ref{fig:urluser} (b) shows the frequency distribution of the difference of the years between citation peak and retweet peak of papers with citations count $\geq$ 20 and retweet count $\geq$ 10 (1881 papers). Higher thresholds are taken to showcase results for dense distributions.
Interestingly, the positive difference suggests that the citation peak occurs after the retweet peak for the majority of the papers. Table \ref{tab:table5} shows a high positive correlation between the attributes at the peak difference of two years. This experiment opens avenues for analysis of temporal peaks in causal relations of social media discourse.




\subsection{Links and Users of tweets}
\label{subsec:links}
The shortened links present in the tweets posed a challenge while mapping papers with tweets. Except for ``fb'', ``lnkd'' and ``github'', all other domains in the Top-10 most occurring domains are link shortening services supported by Twitter or a third party. These services change the text corresponding to the actual web address and provide a short version of the link which forwards the user to the actual address. For e.g. a paper with ``\textit{2008.01342}'' as arXiv ID is embedded as ``\textit{https://ift.tt/2Dkr4rw}'' in one of the tweets. 

All users except ``\textit{@animesh1977}'' amongst the top-10 users posting unique arXiv papers on Twitter are \textit{automated bots} which regularly crawl arXiv and update new papers on Twitter. This shows that automated bots are very prominent in spreading scholarly documents.

\begin{table}
\caption{Pearson's Correlation values of papers with High Citation ($\geq$20) and High Retweet ($\geq$10). R=Retweet, C=Citation, L=Likes}
\label{tab:table5}
\centering
{\begin{tabular}{|c|c|c|c|c|}
\hline Peak Difference & Paper Count & RC Corr & LC Corr & RL Corr \\\hline
2 & 902 &\textbf{0.236} & \textbf{0.421}& \textbf{0.301}\\ \hline
1 & 431 & 0.079 & 0.072 & 0.056 \\\hline
0 & 678 & -0.058 & -0.037 & 0.042 \\ \hline 
\end{tabular}}
\end{table}


\section{Conclusion and Future Work}
In this work, we introduce a novel method to collect better data for boosting the research in analyzing the relations between social media and scientific literature by exploiting temporal features, metadata of tweets, and citation information. This work provides directions that can act as an entry point to accurately analyze the chaotic behavior of social media in the scholarly ecosystem. In the future, \textit{\dataname{}} can be made more robust by incorporating semantics of tweets, keywords from popular scientific publishers with dense citation data, and tweets from the network of top researchers. Also, works in citation prediction have utilized features related to the metadata of a paper \cite{b1}. However, these prediction models still lack features concentrating on the medium of the spread of information. One can explore the possibility of utilizing \textit{\dataname{}} in predicting the impact.


\end{document}